\documentclass[lettersize,journal]{IEEEtran}
\usepackage[utf8]{inputenc}
\usepackage{graphicx} 
\usepackage{amsmath}
\usepackage[linesnumbered,ruled,vlined]{algorithm2e}
\usepackage{float}
\usepackage{subcaption}
\usepackage{enumerate}
\usepackage{color}
\usepackage[margin=1in]{geometry}
\usepackage{cleveref}
\usepackage{booktabs}
\usepackage{authblk}
\usepackage{cleveref}
\usepackage{acronym}

\crefname{figure}{Fig.}{Figs.}
\Crefname{figure}{Fig.}{Figs.}

\title{Geometric Performance Analysis of Doppler-Based Positioning with a Single LEO Satellite}

\author{}
 \author{Qi~Liu,
Marc~Fernandez-Temprado,
Antoni~Reus-Bergas,
Gonzalo~Seco-Granados,~\IEEEmembership{Fellow,~IEEE},
and Jose~A.~Lopez-Salcedo,~\IEEEmembership{Senior~Member,~IEEE}%
\thanks{This work has been partly supported by the Catalan Government
in the framework of the NewSpace Strategy of Catalonia and by the
Spanish Agency of Research (AEI) under grant PID2023-152820OB-I00
funded by MICIU/AEI/10.13039/501100011033 and by ERDF/EU.
(Corresponding author: Qi Liu.)}%
\thanks{Qi Liu, Marc Fernandez-Temprado, Antoni Reus-Bergas,
Gonzalo Seco-Granados, and Jose A. Lopez-Salcedo are with the
Department of Telecommunications and Systems Engineering,
Universitat Autònoma de Barcelona, 08193 Bellaterra, Barcelona, Spain
(e-mail: Qi.Liu@uab.cat; marc.fernandezt@gmail.com;
Antoni.Reus@uab.cat; Gonzalo.Seco@uab.cat; Jose.Salcedo@uab.cat).}%
}
 
\date{December 2025}

\begin{document}

\maketitle
\begin{abstract}
   Low Earth orbit (LEO) satellites are increasingly considered as complementary signal sources for positioning, navigation, and timing (PNT), particularly for Internet of Things (IoT) receivers operating in GNSS-challenged environments. In such scenarios, low-power receivers may only collect sparse Doppler measurements from a single LEO satellite during a short observation window. This paper analyzes the geometric performance of single-satellite LEO Doppler positioning under these constrained observation conditions. A Doppler-based dilution of precision (DDOP) framework is used to predict directional positioning uncertainty under different satellite-user geometries and sparse observation settings. The framework is evaluated through an Orbcomm-based LEO case study using TLE-propagated satellite trajectories. The analysis considers both system-level geometric factors and receiver-level observation parameters, including the maximum elevation of the satellite pass, the user location relative to the satellite ground track, the number of Doppler measurements, the inter-measurement interval, and the observation window placement. An accumulated observation time analysis is further introduced to separate the error reduction into measurement-count and residual geometric contributions. The results reveal pronounced directional anisotropy in the considered Orbcomm-based single-satellite scenario. The cross-track direction is weakly constrained and can experience severe error amplification, whereas the along-track direction is more strongly constrained by Doppler measurements. The analysis also shows that cross-track improvement is mainly driven by the geometry captured during the informative portion of the satellite pass, rather than by measurement accumulation alone. These findings provide practical guidance for pass selection, observation-window placement, and sparse measurement scheduling for low-power IoT receivers, and offer insights into geometry-aware single-satellite LEO-PNT operation using Orbcomm-like LEO Doppler sources.
\end{abstract}


\section{Introduction}
Positioning, Navigation, and Timing (PNT) services delivered by Global Navigation Satellite Systems (GNSS) have become fundamental to modern society \cite{chen2023enhancing,wang2025multipath}, underpinning a wide range of civilian and industrial applications, from everyday smartphone use \cite{zangenehnejad2021gnss} to precision operations in sectors such as agriculture \cite{guo2018multi} and logistics \cite{jhanjhi2024global}. Moreover, critical infrastructure \cite{falletti2018synchronization} and security operations \cite{griffioen2013suitability} rely on GNSS for time synchronization and navigation, further increasing societal dependence on satellite-based PNT services. In the Internet of Things (IoT) context, reliable positioning is also required by large numbers of low-cost devices used in environmental monitoring, maritime sensing, emergency response, and infrastructure monitoring \cite{li2022tightly}. However, many IoT devices operate under stringent constraints in terms of power consumption, hardware complexity, and signal availability \cite{janssen2023survey}. In addition, GNSS performance can be severely degraded in urban canyons, indoor-like environments, dense vegetation, and scenarios affected by intentional or unintentional interference \cite{ding2023performance}. These limitations motivate the development of complementary PNT solutions that can provide positioning support under GNSS-challenged conditions. 

Among complementary PNT sources, signal of opportunity (SOOP) has attracted increasing attention. Terrestrial signals, such as cellular base stations \cite{razavi2018positioning} and broadcast transmitters \cite{li2022performance}, can support positioning in populated areas, but their availability is limited in remote, maritime, or infrastructure-sparse regions. In contrast, Low Earth Orbit (LEO) satellites provide global or near-global coverage and move rapidly with respect to terrestrial users \cite{allahvirdi2022precise,janssen2023survey}. This rapid motion induces strong Doppler variations, that can be exploited for positioning even when the transmitted signals are not originally designed for navigation \cite{stock2023leo}. These features make LEO satellites a promising complementary PNT source for IoT devices operating in GNSS-challenged environments or in areas where terrestrial infrastructure is unavailable \cite{ji2023opportunistic,eissfeller2024comparative}.

The advent of the New Space economy has accelerated the deployment of LEO constellations and has motivated researchers to exploit existing LEO signals for PNT applications \cite{ge2020leo,ccelikbilek2025optimization}. Recent studies have investigated error sources and processing methods in LEO-PNT systems \cite{kassas2024leo,khairallah2023ephemeris}, including orbit prediction \cite{ge2020improving}, differential correction \cite{saroufim2024analysis,hasan2024double}, and inertial-aided compensation \cite{saroufim2023simultaneous}. While these efforts have advanced the modeling and mitigation of LEO-PNT error sources, it remains important to understand how satellite-user geometry and receiver-side observation conditions affect the achievable positioning performance, especially when only sparse Doppler measurements are available. This issue is particularly relevant for low-power IoT receivers, for which duty-cycled operation, limited receiver capability, and environmental blockage may restrict signal acquisition to sparse and short-duration observations.

\subsection{Role of Dilution of Precision (DOP)}
As research on LEO-based PNT continues to expand, it is important to consider not only measurement error mitigation and processing techniques, but also the relative geometry between the satellite and the receiver. This geometry plays a fundamental role in determining achievable positioning accuracy \cite{morales2020gdop}. This influence is commonly characterized by the Dilution of Precision (DOP), a metric that quantifies how observation geometry amplifies measurement errors into position-domain errors. For LEO-based positioning, this geometric effect is particularly important because the satellite-receiver geometry evolves rapidly within a single pass due to the high orbital velocity of LEO satellites \cite{mclemore2022dop,ferre2021comparison, reid2018broadband}.

In conventional multi-satellite GNSS positioning, geometric diversity is mainly provided by the spatial distribution of simultaneously visible satellites \cite{morales2020gdop, xue2015positioning}. In contrast, for single-satellite Doppler positioning, geometric diversity is generated by the temporal evolution of the satellite-user line-of-sight (LOS) vector during the observation window. Therefore, the achievable positioning accuracy depends not only on the Doppler measurement noise, but also on how the measurements are distributed in time and where the observation window is located within the satellite pass.

Previous studies have investigated DOP and DDOP analyses for LEO-based positioning systems, including constellation-level comparisons with GNSS \cite{psiaki2021navigation}, generalized Doppler-based accuracy factors \cite{ baron2024implementation}, and DDOP-based constellation optimization \cite{morales2020gdop}. These studies provide useful tools for evaluating LEO positioning geometry. However, most of them focus on constellation-level performance, multi-satellite availability, or general DDOP formulations. Less attention has been given to the receiver-level design questions that arise when only one LEO satellite is available and only sparse Doppler measurements can be collected. In particular, it remains necessary to clarify how how sparse receiver-side observation settings and satellite-user geometric conditions jointly affect the positioning error in the single-satellite case.

\subsection{Motivation for Single-Satellite Positioning}
While large LEO constellations have received significant attention in recent years, small and incremental LEO deployments remain relevant for experimental demonstrations, institutional systems, and early-stage PNT services \cite{roberto25}. In such scenarios, dense multi-satellite visibility cannot always be guaranteed, making positioning under limited LEO satellite availability an important problem for both system design and receiver operation.

Limited satellite availability can become more severe at the receiver level. Environmental blockage and other signal-degraded conditions may reduce the number of visible satellites, while low-power IoT receivers may only operate during short observation windows. As a result, a receiver may obtain only sparse Doppler measurements from a single LEO satellite during one pass. Characterizing this single-satellite case is therefore essential for assessing the feasibility of LEO Doppler positioning under constrained IoT operating conditions.

These factors motivate the present study on single-satellite LEO Doppler positioning. In this problem, positioning performance is affected by two groups of parameters. The first group is related to system-level geometry, including the maximum elevation of the satellite pass and the user position relative to the satellite ground track. These parameters determine the geometric observability of the single-satellite Doppler solution and the directional distribution of the positioning error. The second group is related to receiver-level observation strategies, including the number of Doppler measurements, the inter-measurement interval, and the observation window placement within a satellite pass. These parameters are particularly relevant for low-power IoT receivers, as they determine how sparse measurements are scheduled during a limited observation window.
\subsection{Contributions of This Work}
In this work, the proposed DDOP formulation is intended to be general for single-satellite multi-epoch Doppler positioning. The numerical analyses are conducted using Orbcomm TLE-propagated satellite trajectories as a representative LEO Doppler source. Therefore, the reported numerical values and practical operating points should be interpreted within this Orbcomm-based case study, whereas the underlying geometric interpretation is more broadly applicable to single-satellite LEO Doppler positioning. The main contributions of this paper are as follows.

First, a DDOP-based analytical framework is developed for single-satellite multi-epoch Doppler positioning. The multi-epoch formulation makes it possible to estimate the user position even when only one satellite is available. The framework predicts how Doppler measurement errors propagate into the receiver position estimate and characterizes the resulting along-track and cross-track error components. The analytical prediction is evaluated through Orbcomm-based simulations under different satellite pass geometries.

Second, the impact of system-level satellite-user geometric parameters is investigated, including the maximum elevation of the satellite pass and the user position relative to the satellite ground track. These analyses show how different geometric configurations affect observability and directional error behavior in single-satellite Doppler positioning, thereby supporting geometry-aware pass selection. The results provide insights for both dedicated LEO-PNT systems and opportunistic use of LEO communication satellites for positioning \cite{picchi2025}.

Third, the impact of receiver-level sparse observation parameters is systematically analyzed, including the number of Doppler measurements, the inter-measurement interval, and the observation window placement. After these parameters are examined, the accumulated-observation-time behavior is further analyzed to separate the error reduction into a measurement-count contribution and a residual geometric contribution, showing that longer observation windows improve the solution not only by increasing the number of measurements but also by capturing the informative geometric evolution of the satellite pass.

Finally, the results provide practical guidelines for lightweight LEO-PNT operation, including how to distribute sparse Doppler measurements in time, when to collect measurements during a pass, and which satellite-user geometries should be preferred or avoided. These are key guidelines for small positioning devices in IoT applications, where stringent computational and power limitations must be accounted for.

\section{DDOP-BASED MODELING AND ERROR CHARACTERIZATION}
This section presents the methodology used to evaluate the geometric performance of single-satellite Doppler-based positioning. The Doppler measurement model is first introduced for the single-epoch case and is then extended to a multi-epoch formulation. The unknown receiver state is estimated using nonlinear and weighted least squares. Based on the linearized model, the DDOP is derived to characterize how Doppler measurement errors are amplified by the observation geometry. The DDOP-based covariance is then related to the positioning error and transformed into along-track and cross-track components. Finally, the common simulation setup used in the following analyses is described.

\subsection{Single-Epoch Doppler Shift Signal Model}
Doppler shift measurements at a receiver are obtained from the difference between the instantaneous received carrier frequency and the nominal transmitted carrier frequency. For a given satellite of interest, Doppler shift can be expressed in units of m/s as a range rate $\dot\rho$, as follows:
\begin{eqnarray}
\begin{aligned}
    \dot \rho  (t) &= -\lambda f_D (t) \\
    &= -\mathbf{v}_s (t)^T  \frac{\mathbf{r}(t') - \mathbf{r}_s (t )}{\|\mathbf{r}(t')-\mathbf{r}_s (t )\|} + c\dot\delta_d 
\end{aligned}
\end{eqnarray}
where $\lambda$ is the wavelength in meters, $f_D$ is the Doppler shift in Hz, $t'$ and $t$ denote the signal reception time and signal transmission time, respectively. In turn, $\mathbf{r}_s(t),\mathbf{v}_s(t)$ denote the satellite position and velocity at the transmission time, $\mathbf{r}(t')$ denotes the unknown user's position at the reception time, and $\dot\delta_d$ denotes the clock drift. 

Two-Line Element (TLE) files are widely used to provide publicly available orbital parameters for satellites, allowing their positions to be propagated using models such as the Simplified General Perturbations model 4 (SGP4). However, due to the simplified orbital representation and model limitations, satellite positions computed from TLE files are subject to certain errors that can be decomposed into along-track and cross-track components. The along-track errors can be partly compensated by introducing a time offset $\delta_c$ \cite{guo2025burst} on the time at which the satellite velocity and position are computed, as indicated in Eq. (\ref{eqDopSigModel}). This equation assumes as well, for the sake of simplicity, that the user is static and thus the time dependence on $\mathbf{r}(t)$ can be dropped hereafter. This leads to 
\begin{eqnarray}
    \dot \rho (t) &=& -\mathbf{v}_s(t - \delta_c)^T \frac{\mathbf{r} - \mathbf{r}_s(t - \delta_c)}{\|\mathbf{r}-\mathbf{r}_s(t-\delta_c)\|} + c\dot\delta_d\label{eqDopSigModel}\\
    &=&h(t;\boldsymbol{\theta})
\end{eqnarray}
This model can be written compactly as a nonlinear function $h(t,\boldsymbol{\theta})$ that relates the set of unknowns $\boldsymbol{\theta}=[\mathbf{r}^T,\dot\delta_d ,\delta_c]^T$ with the observed Doppler shift $\dot\rho(t)$. Finally, if we incorporate the measurement noise $\epsilon(t)$, the single-satellite single-epoch Doppler shift measurement $z(t)$ is written as
\begin{equation}
z(t)=h(t;\boldsymbol{\theta})+\epsilon(t).\label{eqDopShift}
\end{equation}

\subsection{Multi-Epoch Doppler Shift Signal Model}

Unlike conventional GNSS positioning, where multiple satellites are usually observed simultaneously, a LEO-based positioning scenario may provide only one observable satellite at a given epoch. Therefore, positioning with a single LEO satellite must rely on a multi-epoch formulation, in which Doppler measurements are accumulated as the satellite moves along its pass. This can easily be done by collecting a set of $M$ measurements and stacking them into an ($M\times 1$) vector $\mathbf{z}=[\dot\rho(t_1),\dot\rho(t_2),\ldots,\dot\rho(t_M)]^T$, so that Eq. (\ref{eqDopShift}) becomes
\begin{equation}
\mathbf{z}=\mathbf{h}(\boldsymbol{\theta})+\boldsymbol\epsilon\label{eqDefZ}
\end{equation}
where $\mathbf{h}(\boldsymbol{\theta})=[h(t_1;\boldsymbol{\theta}),h(t_2;\boldsymbol{\theta}),\ldots,h(t_M;\boldsymbol{\theta})]^T$ is the stacked nonlinear observation function, and $\boldsymbol{\epsilon}=[\epsilon(t_1),\epsilon(t_2),\ldots,\epsilon(t_M)]^T$ is the measurement noise vector.

The stacked function $\mathbf{h(t;\boldsymbol{\theta})}$ contains the temporal geometric information provided by the selected Doppler measurement epochs. In the single-satellite case, this information is generated by the time variation of the satellite-user line-of-sight vector during the observation window. Therefore, the achievable positioning accuracy depends strongly on which epochs are included in $\mathbf{z}$. This geometry-driven error amplification is evaluated using a Doppler-based dilution of precision (DDOP), which is introduced in the following subsection.

\subsection{Nonlinear Least Squares (NLS) Estimation\label{section:NLS}}
Since the observation model in (\ref{eqDefZ}) is nonlinear with respect to the unknowns, the state cannot be obtained directly in closed form. In order to circumvent this limitation, a first-order Taylor linearization of the observation function near some initial value $\boldsymbol{\theta}_0$ is applied. This leads to
\begin{equation}
\mathbf{h}(\boldsymbol{\theta}) \approx \mathbf{h}(\boldsymbol{\theta}_0) + \left( \frac{\partial \mathbf{h}(\boldsymbol{\theta}) }{\partial \boldsymbol{\theta}}\right) \bigg|_{\boldsymbol{\theta}=\boldsymbol{\theta}_0}  \Delta\boldsymbol{\theta}       
\end{equation}
where $\Delta\boldsymbol{\theta}=\boldsymbol{\theta}-\boldsymbol{\theta}_0=[\Delta r_x,\Delta r_y,\Delta r_z,\Delta\dot{\delta}_d ,\Delta \delta_c ]^T$ denotes the state correction with respect to the linearization point. In this way, the Doppler measurements in Eq. (\ref{eqDefZ}) can be expressed as
\begin{equation}
\mathbf{z}\approx\mathbf{h}(\boldsymbol{\theta}_0) + \left( \frac{\partial \mathbf{h}(\boldsymbol{\theta}) }{\partial \boldsymbol{\theta}}\right) \bigg|_{\boldsymbol{\theta}=\boldsymbol{\theta}_0}  \Delta\boldsymbol{\theta}+\boldsymbol{\epsilon}
\end{equation}
The residual between the measured and modeled Doppler observations can then be written as,
\begin{equation}
\Delta\mathbf{z}(\boldsymbol{\theta}_0) \approx \mathbf{H}(\boldsymbol{\theta}_0)  \Delta\boldsymbol{\theta}+\boldsymbol{\epsilon}\label{eqRes}
\end{equation}
where $\Delta\mathbf{z}(\boldsymbol{\theta}_0)=\mathbf{z}-\mathbf{h}(\boldsymbol{\theta}_0)$ and the Jacobian matrix $\mathbf{H}(\boldsymbol{\theta}_0)=(\partial\mathbf{h}(\boldsymbol{\theta})/\partial\boldsymbol{\theta})_{|\boldsymbol{\theta}=\boldsymbol{\theta}_0}$ becomes
\begin{equation}
    \mathbf{H}(\boldsymbol{\theta}_0)=
\begin{bmatrix}
\frac{\partial \dot \rho (t_{1}) }{\partial r_x} & \frac{\partial \dot \rho (t_{1}) }{\partial r_y} &\frac{\partial \dot \rho (t_{1}) }{\partial r_z} & c & \frac{\partial \dot \rho (t_{1}) }{\partial \delta _c}  \\
\frac{\partial \dot \rho (t_{2}) }{\partial r_x} & \frac{\partial \dot \rho (t_{2}) }{\partial r_y} &\frac{\partial \dot \rho(t_{2}) }{\partial r_z} & c & \frac{\partial \dot \rho  (t_{2}) }{\partial \delta _c }  \\
\vdots & \vdots & \vdots\\
\frac{\partial \dot \rho  (t_{M}) }{\partial r_x} & \frac{\partial \dot \rho  (t_{M}) }{\partial r_y} &\frac{\partial \dot \rho  (t_{M}) }{\partial r_z} & c & \frac{\partial \dot \rho  (t_{M}) }{\partial \delta _c }  \\ 
\end{bmatrix}_{\big|\boldsymbol{\theta}=\boldsymbol{\theta}_0}
\end{equation}
whose elements are provided, for the sake of completeness, in Appendix \ref{appA}.

The Jacobian matrix plays a central role in the process of estimating the unknowns, since it characterizes how the measurements vary with respect to the unknown states while linking the geometry to the estimator performance.

Subsequently, instead of directly solving for $\Delta\boldsymbol{\theta}$ in (\ref{eqRes}) using a least squares on the residuals, we will incorporate the fact that measurements at low elevation angles are more affected by noise than those at high elevation angles. In this way, $\Delta\boldsymbol{\theta}$ can be estimated as the solution to the following Weighted Least Squares (WLS) cost function
\begin{equation}
    J(\Delta \boldsymbol{\theta}) = (\Delta \mathbf{z}(\boldsymbol{\theta}_0) - \mathbf{H}(\boldsymbol{\theta}_0)\Delta\boldsymbol{\theta})^\mathrm{T}\mathbf{W}(\Delta \mathbf{z}(\boldsymbol{\theta}_0) - \mathbf{H}(\boldsymbol{\theta}_0)\Delta\boldsymbol{\theta})
\end{equation}
where $\mathbf{W} = \mathrm{diag}
(\frac{1}{\sigma_1^2},\frac{1}{\sigma_2^2},...,\frac{1}{\sigma_M^2})$ is a weighted matrix, adopting an elevation-based weighted model $\sigma^2_i = \frac{1}{\mathrm{sin}^2(E_i)}$ and $E_i$ is the elevation of the single satellite under analysis at the $i$-th measurement. This cost function measures the discrepancy between the model's predicted values and actual observations. Its minimization leads to the following estimate
\begin{equation}
    \Delta\hat{\boldsymbol{\theta}} = \big(\mathbf{H}^T(\boldsymbol{\theta}_0)\mathbf{W}\mathbf{H}(\boldsymbol{\theta}_0)\big)^{-1}\mathbf{H}^T(\boldsymbol{\theta}_0)\mathbf{W}\Delta\mathbf{z}(\boldsymbol{\theta}_0)\label{eqWLSsol}
\end{equation}
which can be implemented in an iterative manner as
\begin{equation}
    \Delta\hat{\boldsymbol{\theta}}_{k} = \big(\mathbf{H}^T(\boldsymbol{\theta}_k)\mathbf{W}\mathbf{H}(\boldsymbol{\theta}_k)\big)^{-1}\mathbf{H}^T(\boldsymbol{\theta}_k)\mathbf{W}\Delta\mathbf{z}(\boldsymbol{\theta}_k).
\end{equation}
then updating the unknown as
\begin{equation}
\hat{\boldsymbol{\theta}}_{k+1}=\hat{\boldsymbol{\theta}}_k+\Delta\hat{\boldsymbol{\theta}}_k
\end{equation}
and the residuals as $\Delta\mathbf{z}(\boldsymbol{\theta}_{k+1})=\mathbf{z}-\mathbf{h}(\boldsymbol{\theta}_{k+1})$, until convergence is achieved on $\hat{\boldsymbol{\theta}}_{k+1}$.

\subsection{DDOP Formulation and Computation\label{section:DDOP}}
DOP characterizes how the observation geometry maps measurement errors into the estimated states \cite{psiaki2021navigation}. For DDOP, the estimated states have different physical units, including position, clock drift, and satellite time offset. Therefore, a scaling operation is required before computing the DOP-like metric from the Doppler measurement matrix.

The products between the Jacobian columns and the corresponding state errors must be consistent with the units of the Doppler-derived range-rate measurements. Therefore, the columns of the Jacobian are scaled using a diagonal scaling matrix $\mathbf{S}$ Eq. (\ref{eqRes}) can then be rewritten as

\begin{eqnarray}
\Delta\mathbf{z}(\boldsymbol{\theta}_0)&\approx&\mathbf{H}(\boldsymbol{\theta}_0)\mathbf{SS}^{-1}\Delta\boldsymbol{\theta}+\boldsymbol{\epsilon}
\\&=&\tilde{\mathbf{H}}(\boldsymbol{\theta}_0)\Delta\tilde{\boldsymbol{\theta}}+\boldsymbol{\epsilon}
\end{eqnarray}
where the scaled Jacobian matrix $\tilde{\mathbf{H}}(\boldsymbol{\theta}_0) = \mathbf{H}(\boldsymbol{\theta}_0)\mathbf{S}$ turns out to be given by
\begin{equation}
\begin{aligned}
&\tilde{\mathbf{H}} = \\[-0.1em]
&\resizebox{1\columnwidth}{!}{$
\begin{bmatrix}
\frac{\partial \dot \rho  (t_{1}) }{\partial r_x}/\gamma &
\frac{\partial \dot \rho  (t_{1}) }{\partial r_y}/\gamma &
\frac{\partial \dot \rho  (t_{1}) }{\partial r_z}/\gamma &
c &
\frac{\partial \dot \rho  (t_{1}) }{\partial\delta_c }/\eta  \\
\frac{\partial \dot \rho  (t_{2}) }{\partial r_x}/\gamma &
\frac{\partial \dot \rho  (t_{2}) }{\partial r_y}/\gamma &
\frac{\partial \dot \rho  (t_{2}) }{\partial r_z}/\gamma &
c &
\frac{\partial \dot \rho  (t_{2}) }{\partial\delta _c }/\eta  \\
\vdots & \vdots & \vdots & \vdots & \vdots \\
\frac{\partial \dot \rho  (t_{M}) }{\partial r_x}/\gamma &
\frac{\partial \dot \rho  (t_{M}) }{\partial r_y}/\gamma &
\frac{\partial \dot \rho  (t_{M}) }{\partial r_z}/\gamma &
c &
\frac{\partial \dot \rho  (t_{M}) }{\partial\delta _c }/\eta
\end{bmatrix}.
$}
\end{aligned}
\end{equation}
Similarly, the scaled residual unknowns $\Delta\tilde{\boldsymbol{\theta}}=\mathbf{S}^{-1}\Delta\boldsymbol{\theta}$ are given by
\begin{equation}
\Delta\tilde{\boldsymbol{\theta}}=\begin{bmatrix}
\gamma\Delta r_x \\
\gamma\Delta r_y \\
\gamma\Delta r_z \\
c\Delta \dot{\delta}_d \\
\eta \Delta\delta_c 
\end{bmatrix}
\end{equation}
with $\gamma$ and $\eta$, scaling factors given by \cite{baron2024implementation}
\begin{eqnarray}
\gamma&=& \left(\frac{1}{1-r_e/a_{orb}}\right)  \sqrt{\frac{\mu}{a_{orb}^3}}\label{gamma}\\ 
\eta &=& \left(\frac{r_e / a_{orb}}{1-r_e/a_{orb}}\right)\left(\frac{\mu}{a_{orb}^2}\right)
\end{eqnarray}
where $r_e$ is the radius of earth 6371 km, $a_{orb}$ is the semi-major axis of the satellite’s orbit and $\mu$ is the Earth’s gravitational constant $3.986004418\cdot 10^{14}$  $m^3/s^2$.

According to the aforementioned formulation, the covariance matrix of the scaled unknown vector $\Delta\tilde{\boldsymbol{\theta}}$ in (\ref{eqWLSsol}) is known to be given by \cite{kay93}
\begin{eqnarray}
\mathbf{C}_{\Delta\hat{\boldsymbol{\theta}}}&=&\mathrm{E}\left[(\Delta\hat{\boldsymbol{\theta}}-\mathrm{E}\left[\Delta\hat{\boldsymbol{\theta}}\right])(\Delta\hat{\boldsymbol{\theta}}-\mathrm{E}\left[\Delta\hat{\boldsymbol{\theta}}\right])^T\right]\\&=&\tilde{(\mathbf{H}}^T\mathbf{W} \tilde{\mathbf{H}})^{-1}.
\end{eqnarray}
In this way, the DDOP is calculated as follows
\begin{eqnarray}
\mathrm{PDDOP} & =& \sqrt{\mathrm{Tr}(\left[\mathbf{C}_{\Delta\hat{\boldsymbol{\theta}}}\right]_{1:3 , 1:3})}\label{eqPDDOP}\\
\mathrm{HDDOP} & =& \sqrt{\mathrm{Tr}(\left[\mathbf{C}_{\Delta\hat{\boldsymbol{\theta}}}\right]_{1:2 , 1:2})}\label{eqHDDOP}\\
\mathrm{CDDOP} & =& \sqrt{\mathrm{Tr}(\left[\mathbf{C}_{\Delta\hat{\boldsymbol{\theta}}}\right]_{4,4})}\label{eqHDDOP}\\
\mathrm{TDDOP} & =& \sqrt{\mathrm{Tr}(\left[\mathbf{C}_{\Delta\hat{\boldsymbol{\theta}}}\right]_{5,5}}\label{eqTDDOP}
\end{eqnarray}
where $\left[\mathbf{C}_{\Delta\hat{\boldsymbol{\theta}}}\right]_{a:b,c:d}$ indicates rows $a$ to $b$ and columns $c$ to $d$ of matrix $\left[\mathbf{C}_{\Delta\hat{\boldsymbol{\theta}}}\right]$. In turn, PDDOP stands for the {\em position} DDOP, HDDOP the {\em horizontal} position DDOP, CDDOP the {\em clock drift} DDOP and TDDOP the {\em time} offset DDOP.

\begin{table}[H]
    \centering
    \caption{DDOP Expected Accuracy and Dimensional Precision Metrics}
    \label{DDOP-metrics}
    \begin{tabular}{cl}
        \hline
        Estimated Quantity & Dimensional Precision Metric \\
        \hline
        $\mathbf{r}$          & $\mathrm{PDDOP}\times\left(\lambda\sigma_{\mathrm{Dopp}} / \gamma\right)$ \\[2pt]
        $\delta_c $      & $\mathrm{TDDOP}\times\left(\lambda\sigma_{\mathrm{Dopp}} / \eta\right)$ \\[2pt]
        $\dot{\delta}_d $ & $\mathrm{CDDOP}\times\left(\lambda\sigma_{\mathrm{Dopp}} / c\right)$ \\
        \hline
    \end{tabular}  
\end{table}

Considering that user location variations predominantly occur within the horizontal plane, we henceforth assume for simplicity that the coordinate in the $z$ direction is known and remains constant, which means that the unknown parameters are $\boldsymbol{\theta}=[r_x,r_y,\dot\delta_d ,\delta_c ]^T$. 

\subsection{DDOP-Based Positioning Error Characterization\label{section:ellipses}}
After the DDOP has been formulated, it is necessary to relate it to the positioning errors used in the following analyses. The DDOP describes how Doppler measurement noise is amplified by the observation geometry. Therefore, the dimensional position-domain covariance can be obtained by combining the scaled covariance matrix with the Doppler noise level and the corresponding scaling factor. For the position-related states, it is given by
\begin{equation}
    \mathbf{C}_{\mathbf{r}} = \left(\frac{\lambda\sigma_D}{\gamma}\right)^2 \left[\mathbf{C}_{\Delta\hat{\boldsymbol{\theta}}}\right]_{1:3 , 1:3}
\end{equation}
where $\sigma_D$ is the standard deviation of the Doppler measurement noise and $\gamma$ is the position scaling factor computed in \cref{gamma}. Since the vertical coordinate is assumed to be known, the corresponding covariance entry is set to zero in the following horizontal error analysis.
\begin{figure}[t]
    \centering
    \includegraphics[width=0.9\linewidth]{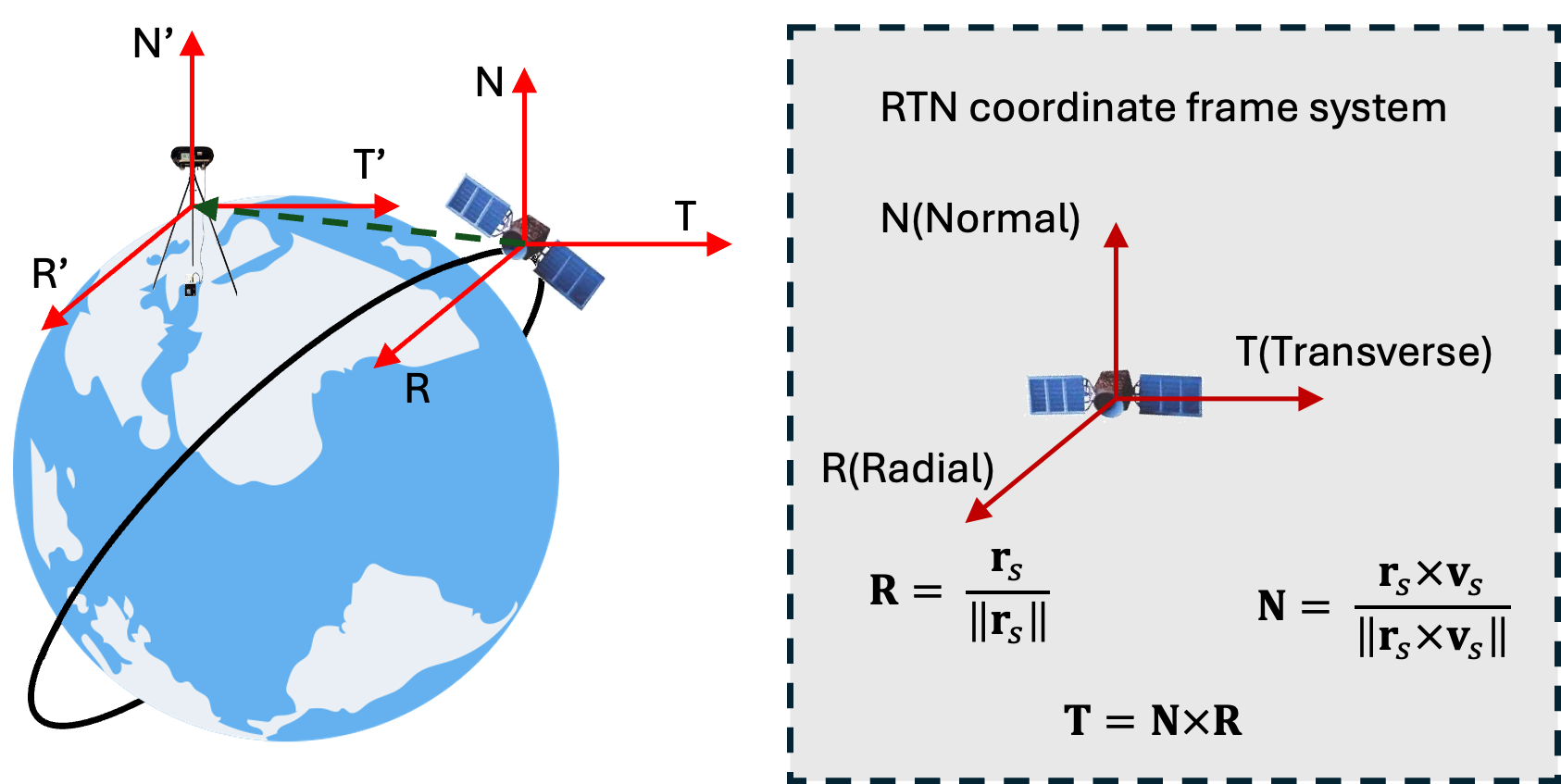}
    \caption{Schematic diagram of the RTN coordinate frame.}
    \label{fig:RTN}
\end{figure}

\begin{figure*}[t]
    \centering
    \begin{subfigure}{0.49\textwidth}
        \centering
        \includegraphics[width=\linewidth]{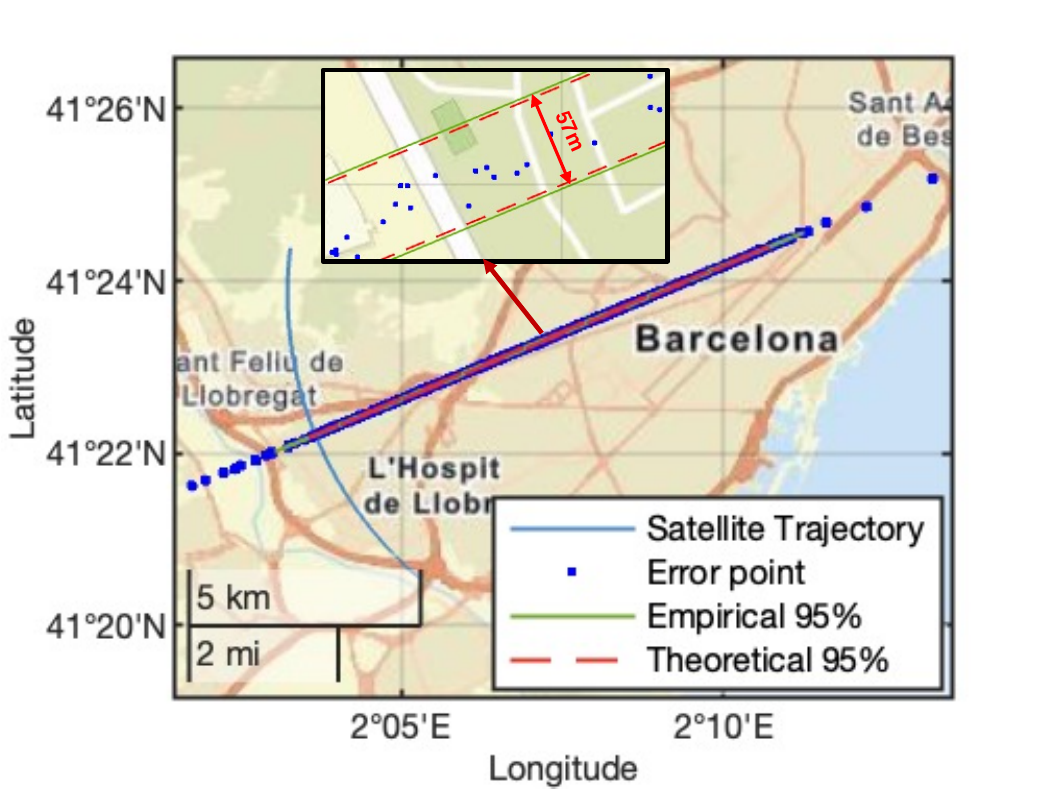}
        \caption{}
        \label{fig:match_theo2sim_a}
    \end{subfigure}
    \hfill
    \begin{subfigure}{0.49\textwidth}
        \centering
        \includegraphics[width=\linewidth]{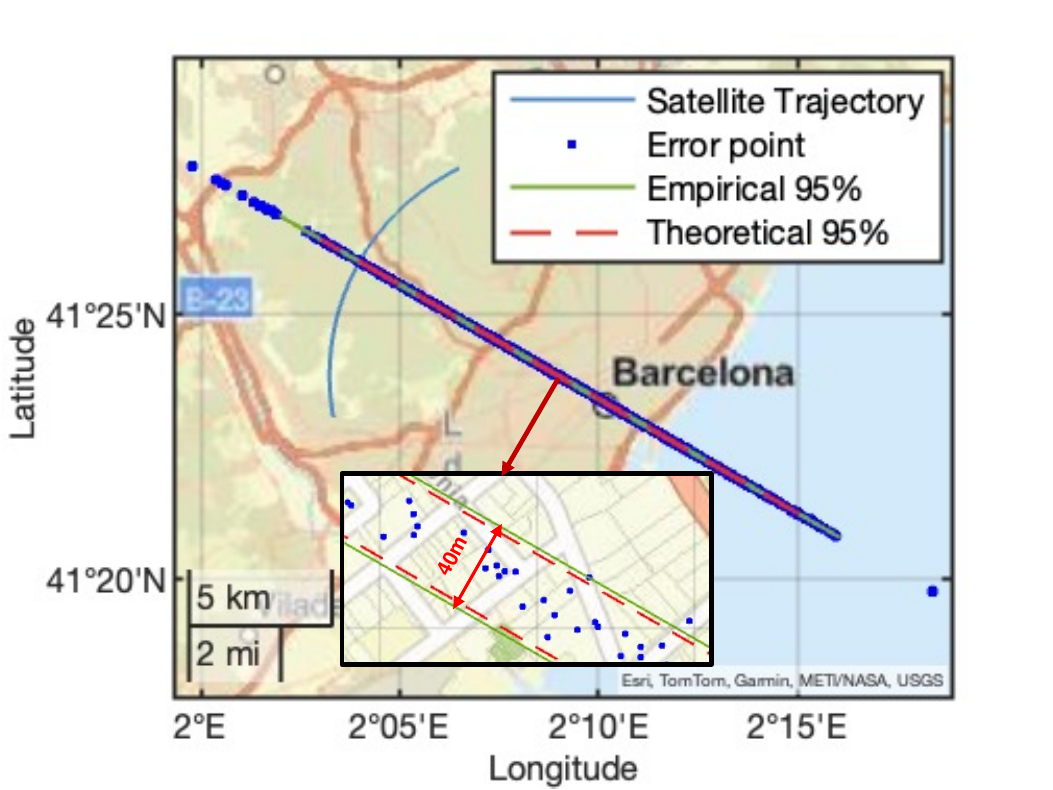}
        \caption{}
        \label{fig:match_theo2sim_b}
    \end{subfigure}
    \caption{Simulation error ellipse and theoretical ellipse of Orbcomm under two different passes.}
    \label{fig:match_theo2sim}
\end{figure*}

For each Monte Carlo simulation result, the positioning error is $\Delta\mathbf{r}$. To characterize the directional behavior of the positioning error, the error vector is expressed in the Radial-Transverse-Normal (RTN) coordinate frame associated with the considered satellite pass, as illustrated in \Cref{fig:RTN}. The RTN frame is constructed using the satellite position $\mathbf{r}_s$ and velocity $\mathbf{v}_s$. In this coordinate frame, the T-axis points along the satellite’s direction of motion (i.e., the along-track direction), while the N-axis corresponds to the cross-track direction. Accordingly, the corresponding unit vectors are defined as
\begin{equation}
    \mathbf{R} = \frac{\mathbf{r}_s}{||\mathbf{r}_s||},\quad \mathbf{N} = \frac{\mathbf{r}_s \times \mathbf{v}_s }{||\mathbf{r}_s \times \mathbf{v}_s||},\quad \mathbf{T} = \mathbf{N} \times \mathbf{R}
\end{equation}
The rotation matrix is then written as $\mathbf{R}_{\mathrm{RTN}} = [\mathbf{R} \quad \mathbf{T} \quad\mathbf{N}]$. Therefore, the positioning error expressed in the RTN frame is obtained as $\Delta\mathbf{r}_{\mathrm{RTN}} =\mathbf{R}_{\mathrm{RTN}}^T \Delta\mathbf{r}$. In this work, the transverse component is referred to as the along-track error $\Delta\mathbf{r}_{\mathrm{RTN}}(2)$, while the normal component is referred to as the cross-track error $\Delta\mathbf{r}_{\mathrm{RTN}}(3)$.

For the validation analysis, 95\% confidence ellipses are constructed from the empirical and theoretical covariance matrices. The empirical covariance is computed from the Monte Carlo positioning errors, whereas the theoretical covariance is obtained from the DDOP-based error covariance. The ellipse orientation and principal axes are determined by the eigenvectors and eigenvalues of the covariance matrix, and the axes are scaled using $\chi^2_{2,0.95} = 5.991$.
\subsection{Simulation Setup \label{section:setup}}
This subsection describes the common simulation setup in our experiments. The receiver is assumed to be located at $[41.3976^\circ, 2.1497^{\circ},60 \mathrm{m}]$, corresponding to Barcelona, Spain. Simulated Doppler observations are generated at a 1 s temporal resolution. The satellite trajectory is propagated from the TLE data of Orbcomm. Different single-satellite pass geometries are then considered depending on the specific experiment. Unless otherwise stated, all experiments are conducted using the simulation setup described above. Doppler measurements are generated from the noise-free model in (\ref{eqDopSigModel}) and corrupted with zero-mean Gaussian noise as in (\ref{eqDopShift}). The receiver state is estimated using the WLS algorithm described in \Cref{section:NLS}.

\begin{figure}[t]
    \centering  
    \includegraphics[width=1\linewidth]{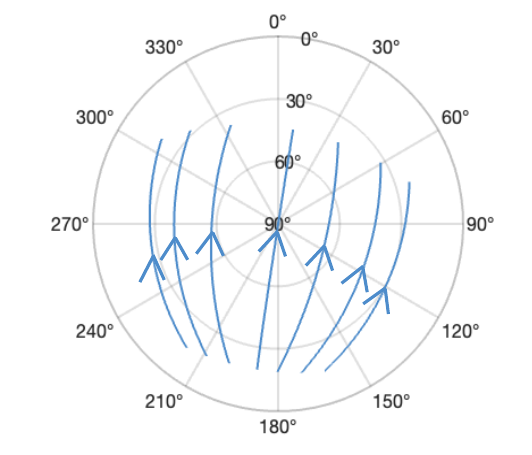}
    \caption{Skyplot for representative satellite pass geometries (arrows indicate the direction of motion).}
    \label{fig:skyplot_inclinations}
\end{figure}

\section{System-Level Geometric Analysis}
This section analyzes the system-level geometry effects in single-satellite LEO Doppler positioning. Before examining individual geometric factors, the DDOP-based positioning error model derived in \cref{section:DDOP} is validated using Monte Carlo simulations. The analysis then focuses on two system-level factors: the maximum elevation of the satellite pass and the user position relative to the satellite ground track. These factors are determined by the satellite-user geometry and are therefore treated separately from receiver-level observation parameters.

\subsection{Validation of the DDOP-Based Positioning Error Model}
To validate the theoretical DDOP obtained in (\ref{eqPDDOP})-(\ref{eqTDDOP}), the values obtained will be compared to the positioning errors incurred when numerically solving the user's position using the iterative WLS framework described in Section \ref{section:NLS}. To do so, positioning errors over 1000 Monte Carlo simulations will be computed, using two different satellite orbits in order to have some diversity on the impact of the actual satellite pass.
    

The comparison between the actual single-satellite positioning errors and the theoretical bounding provided by the DDOP is shown in \Cref{fig:match_theo2sim} for two different satellite passes. In each figure, the ground projection of the passes is indicated by a light blue solid line. The blue dots represent positioning error of each individual simulation with respect to the reference position. To describe the variation patterns of errors in different directions, the horizontal error is decomposed into along-track and cross-track components as defined in \Cref{fig:RTN}.
 \begin{figure}[t]
    \centering  
    \includegraphics[width=1\linewidth]{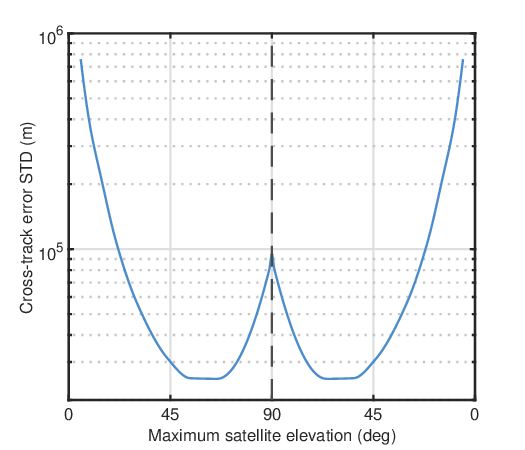}
    \caption{Effect of satellite pass geometry on cross-track error for single-satellite LEO positioning.}
    \label{fig:Inclination_cross}
\end{figure}
The empirical 95\% confidence ellipse (solid green) computed from the positioning errors obtained from the simulations, is compared with the corresponding theoretical 95\% ellipse (red dashed) predicted by the DDOP computation. Both ellipses are constructed according to the procedure described in \Cref{section:ellipses}. For the first satellite pass shown in \Cref{fig:match_theo2sim}(a), the simulated along-track and cross-track errors standard deviations are 28 m and 9932 m, respectively, while their corresponding theoretical values are 27 m and 9629 m, respectively. For the second satellite pass shown in \Cref{fig:match_theo2sim}(b), the simulated along-track and cross-track errors are 20 m and 5561 m, respectively, while their theoretical counterparts are 19 m and 5384 m, respectively. Even though the magnitude of the positioning error varies across different satellite trajectories, the estimated empirical 95\% confidence ellipse and the theoretical 95\% ellipse turn out to be in high agreement. This confirms that the DDOP-based model captures both the dominant error magnitude and the anisotropic direction of the single-satellite Doppler solution.

After this validation, the subsequent analyses use the DDOP-predicted positioning accuracy to evaluate different geometric configurations efficiently.

 \subsection{Impact of Satellite Pass Maximum Elevation}
The geometry of a satellite pass determines the ground-track direction and the temporal evolution of the LOS vector between the satellite and the receiver. Intuitively, different pass geometries lead to different sky-track shapes and LOS variation patterns during a satellite pass, which can significantly affect the performance of single-satellite Doppler positioning. In this subsection, the satellite pass geometry is represented by the maximum elevation angle during the pass, which reflects how close the satellite trajectory passes to the receiver. This quantity should not be confused with the classical orbital inclination defined in TLE files.

 \begin{figure}[t]
    \centering  
    \includegraphics[width=1\linewidth]{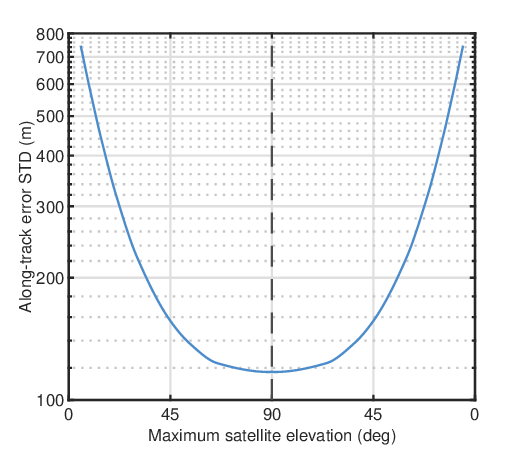}
    \caption{Effect of satellite pass geometry on along-track error for single-satellite LEO positioning.}
    \label{fig:Inclination_along}
\end{figure}

To investigate this effect, we analyze the positioning performance under different satellite pass maximum elevations, as illustrated in \Cref{fig:skyplot_inclinations}. In the experiment, the maximum elevation angle of the satellite pass is varied while keeping the observation duration and measurement conditions unchanged. This controlled setup allows the resulting positioning errors to reveal how different pass geometries influence the positioning performance of a single LEO satellite.

\Cref{fig:Inclination_cross} shows the cross-track positioning error as a function of the satellite pass maximum elevation. A pronounced peak is observed at a maximum elevation angle of $90^\circ$, corresponding to the satellite trajectory passing directly overhead the receiver. In this configuration, the LOS vectors remain nearly coplanar throughout the satellite pass, leading to limited geometric diversity in the cross-track direction. As a result, cross-track observability is significantly degraded, with the estimation error nearly reaching $10^5$ m. In contrast, when the maximum elevation decreases from  $90^\circ$, the cross-track error initially decreases due to the enhanced geometric diversity introduced by the lateral offset of the satellite trajectory. However, when the pass moves further away from the receiver and the maximum elevation becomes too low, the error gradually increases again, as the degradation in overall measurement geometry and the associated increase in DDOP outweigh the benefits brought by additional cross-track information. This results in a non-monotonic and approximately symmetric error profile, indicating that an optimal maximum elevation angle exists at which the cross-track positioning error is minimized.

\Cref{fig:Inclination_along} illustrates the variation pattern of the along-track error, which exhibits an opposite trend compared to the cross-track error. The along-track error reaches its minimum, approximately on the order of $10^2$ m, when the maximum satellite elevation approaches $90^\circ$. As the pass deviates from this near-overhead configuration, along-track observability diminishes, causing the estimation error to increase continuously.

The results indicate that the satellite pass maximum elevation determines a clear trade-off between along-track and cross-track observability. A near-zenith pass improves along-track estimation because the Doppler measurements change more strongly over time, but it also weakens cross-track estimation because the observation geometry remains nearly coplanar. This phenomenon highlights a fundamental limitation of single-satellite positioning using Doppler measurements.

Combining the results in \Cref{fig:Inclination_cross,fig:Inclination_along}, the best overall single-satellite positioning performance in the considered scenario appears when the maximum satellite elevation is around $60^\circ$. At this point, the cross-track error, which largely dominates in the overall positioning error, reaches its minimum, while the along-track error remains close to its minimum value, indicating a near-optimal trade-off between the two directions.
 \begin{figure}[t]
    \centering
    \begin{subfigure}{1\linewidth}
      \centering
      \includegraphics[width=1\linewidth]{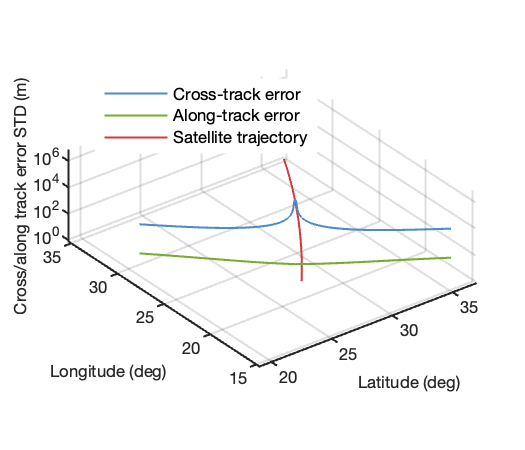}
      \caption{}
    \end{subfigure}
    
    \begin{subfigure}{1\linewidth}
      \centering
      \includegraphics[width=1\linewidth]{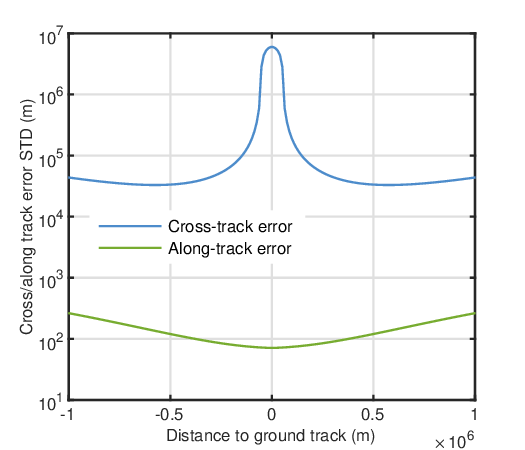}
      \caption{}
    \end{subfigure}
    \caption{Variation of positioning accuracy with ground-track distance for users distributed along the cross-track direction.}
    \label{fig:Multi_pos}
\end{figure}
\subsection{Impact of User Position Relative to the Satellite Pass}
Positioning accuracy fundamentally depends on the relative geometry between the satellite and the user, whose effect is analyzed in this subsection for a  fixed orbital configuration. The first half of this subsection analyzes the variation patterns of positioning errors when the user's positions are distributed along the cross-track direction, as this corresponds to the weakest geometric constraint in a single-satellite LEO system.

As shown in \Cref{fig:Multi_pos} (a), the horizontal plane represents the user latitude and longitude, while the vertical axis represents the directional positioning error at each user location. The visualization shows the satellite ground track together with the along-track and cross-track error variations for users distributed across the track. The error surface exhibits strong directional dependence, indicating that the positioning accuracy is highly sensitive to the user position relative to the satellite ground track.

\begin{figure}[t]
    \centering
    \begin{subfigure}{1\linewidth}
      \centering
      \includegraphics[width=1\linewidth]{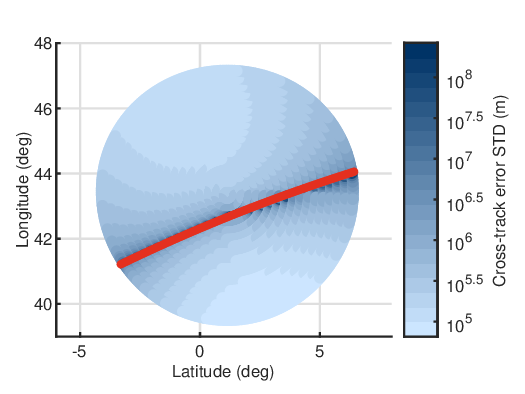}
      \caption{}
    \end{subfigure}
    
    \begin{subfigure}{1\linewidth}
      \centering
      \includegraphics[width=1\linewidth]{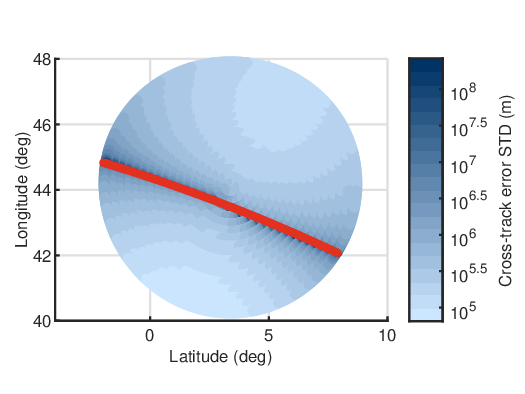}
      \caption{}
    \end{subfigure}
    \caption{Spatial distribution of cross-track error in single-satellite LEO positioning systems under different satellite passes.}
    \label{fig:allpos_cross}
\end{figure}

 \begin{figure}[t]
    \centering
    \begin{subfigure}{1\linewidth}
      \centering
      \includegraphics[width=\linewidth]{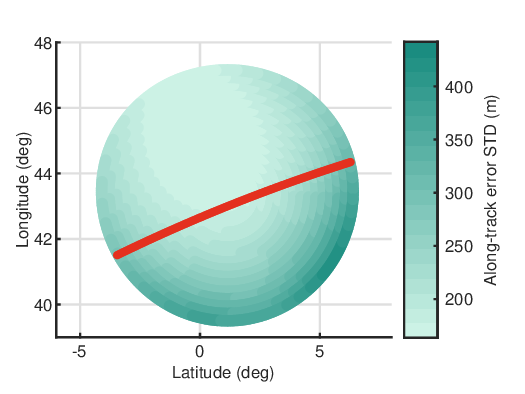}
      \caption{}
    \end{subfigure}
    
    \begin{subfigure}{1\linewidth}
      \centering
      \includegraphics[width=1\linewidth]{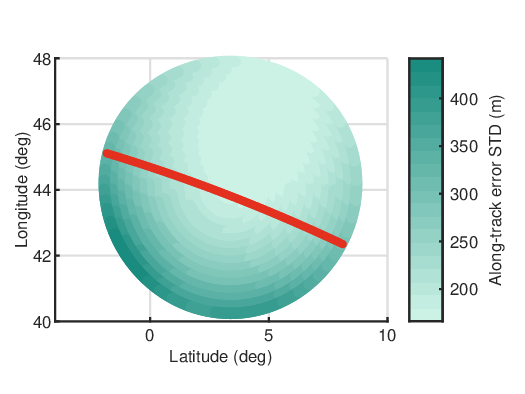}
      \caption{}
    \end{subfigure}
    \caption{Spatial distribution of along-track error in single-satellite LEO positioning systems under different satellite passes.}
    \label{fig:allpos_along}
\end{figure}

\Cref{fig:Multi_pos} (b) further quantifies this phenomenon by illustrating how the error distribution varies with the user's distance from the ground track. Near the ground track, the cross-track error exhibits a pronounced peak, which can be attributed to the degraded geometric observability in the cross-track direction. As the user moves away from the ground track, the lateral offset improves the observation geometry, and the cross-track error decreases accordingly.

In contrast, the along-track error varies more smoothly with the user distance from the ground track. This indicates that the Doppler geometry along the satellite motion direction remains more stable. The difference between the two error components highlights the anisotropic nature of single-satellite Doppler positioning: the cross-track component is strongly affected by the instantaneous satellite-user geometry, whereas the along-track component benefits more directly from the temporal diversity introduced by satellite motion.

To more comprehensively investigate how the positioning error depends on the user location relative to the satellite pass, the analysis is extended to users distributed over the entire observation region. In this analysis, the satellite orbit and all observation settings are kept fixed, so that only the user position varies. This enables a systematic examination of the spatial distribution and directional characteristics of the positioning errors in the single-satellite case.

The cross-track error results in \cref{fig:allpos_cross} indicate that under different passes, high-error zones consistently align with the satellite's ground track. The direction experiencing the most severe error increase corresponds precisely to the direction with weaker cross-track geometric constraints, with error peaks reaching approximately $10^8$ m. Although the high-error zones vary with the satellite's orbit, their magnitude and structural characteristics remain similar across different orbits.
 \begin{figure}[t]
    \centering  
    \includegraphics[width=.99\linewidth]{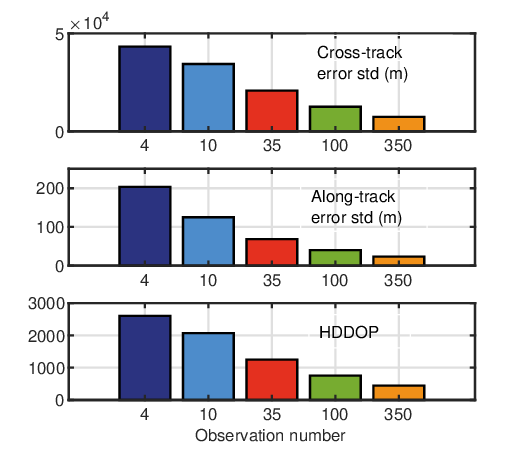}
    \caption{Cross-track error, along-track error and HDDOP under different observation number.}
    \label{fig:Orb108_num}
\end{figure}

In comparison, the along-track error is minimized along the satellite ground track direction in \cref{fig:allpos_along}, where single-satellite Doppler measurements offer the strongest geometric constraint along the satellite's motion direction. Both passes display similar spatial distribution patterns and error levels, with error magnitudes remaining within 500 meters. This indicates that in single-satellite Doppler positioning, the along-track direction yields significantly more precise information from the measurement data.

Overall, the results show that single-satellite Doppler positioning can provide useful directional positioning information under favorable geometric conditions, although its performance remains strongly limited by cross-track observability. The analysis also indicates that the positioning accuracy is influenced by cross-track geometric relationships, particularly when the receiver is located close to the satellite’s ground track. These findings suggest that, while single-satellite observations can already provide useful positioning capability in certain scenarios, further improvements can be achieved by incorporating measurements with more diverse geometries, such as those obtained from multiple passes or complementary trajectories.

\section{Receiver-Level Sparse Observation Analysis}
This section investigates receiver-level sparse observation strategies for single-satellite LEO Doppler positioning. Three observation parameters are first analyzed separately: the number of Doppler measurements, the inter-measurement interval, and the observation window placement within a satellite pass. The accumulated observation time analysis is then used to examine their combined effect during a continuous single-pass observation. This decomposition separates the total positioning improvement into a measurement-count gain and a residual geometric gain, clarifying how measurement accumulation and geometry evolution jointly improve the solution.
 \begin{figure}[t]
    \centering  
\includegraphics[width=1\linewidth]{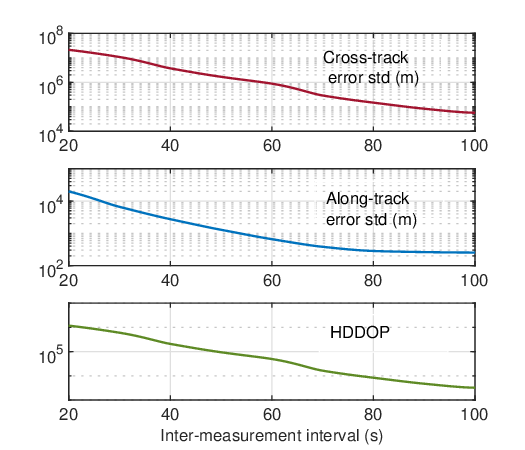}
    \caption{Cross-track error, along-track error and HDDOP under different inter-measurement intervals.}
    \label{fig:Orb108}
\end{figure}
\subsection{Impact of the Number of Doppler Measurements}
The number of Doppler measurements directly affects the geometry and observability of single-satellite positioning. To assess its impact on positioning performance, experimental sub-datasets are constructed from a fixed time interval. This interval corresponds to the Orbcomm satellite pass across the central region of the sky, with a maximum elevation angle of about 70°. Doppler observations are then uniformly generated within this interval according to the prescribed observation quantity. By keeping the trajectory duration fixed, the influence of trajectory-length variations is eliminated, enabling a fair comparison of positioning performance under different observation quantities.

\Cref{fig:Orb108_num} shows the influence of the number of Doppler observations on the positioning performance of the single-satellite solution. As the number of measurements increases from 4 to 350, the cross-track error decreases from approximately $4.2\cdot 10^4$ m to below $10^4$ m. The along-track error also decreases, but its magnitude remains much smaller than that of the cross-track component. The HDDOP follows the same decreasing trend, confirming that the improvement is driven by a stronger Doppler observation geometry. However, the reduction becomes less pronounced as more measurements are added within the fixed observation interval, indicating diminishing returns once the main temporal geometry has been sampled.
\begin{figure}[t]
    \centering
    \includegraphics[width=1\linewidth]{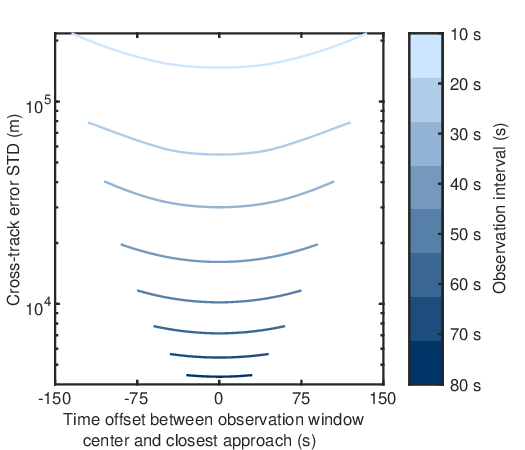}
    \caption{Impact of observation time and interval on cross-track accuracy using four Doppler measurements.}
    \label{fig:crosstrack_startpoint}
\end{figure}
\subsection{ Impact of the Inter-Measurement Interval}

Considering the limited visibility and measurement availability that may occur in practical LEO-PNT scenarios, it is important to understand the performance limits of single-satellite Doppler positioning when only a minimal number of observations can be collected. In the model adopted in this study, four parameters are estimated as unknowns; therefore, a navigation solution can theoretically be obtained with four fully independent Doppler observations. As demonstrated in the previous experiment, positioning performance improves with an increase in the amount of measurements. However, when the number of available observations is constrained, an important question is how these limited measurements should be distributed in time. The time separation between consecutive Doppler measurements, referred to here as the inter-measurement interval, is analyzed in this subsection. To isolate the effect of the inter-measurement interval, the starting point is fixed, corresponding to the instant when the signal was first captured.

\Cref{fig:Orb108} shows the along-track error, cross-track error, and HDDOP as functions of the inter-measurement interval. Both error components decrease as the interval increases, indicating that a larger temporal separation allows the satellite motion to provide more diverse Doppler geometry. The most pronounced improvement occurs at short intervals, approximately between 20 s and 40 s. Beyond this range, the curves continue to decrease but with a slower rate, suggesting that the main geometric diversity has already been captured. The cross-track error remains much larger than the along-track error for all tested intervals, confirming that the cross-track direction is the limiting component in the four-measurement single-satellite solution. The HDDOP follows the same trend as the positioning errors, further indicating that the performance improvement is primarily geometry-driven.

\begin{figure}[t]
    \centering
    \includegraphics[width=1\linewidth]{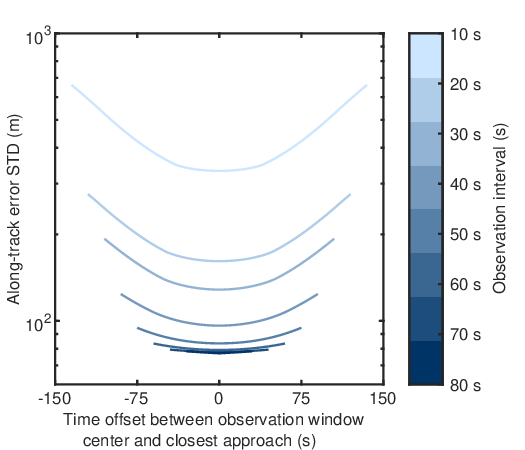}
    \caption{Impact of observation time and interval on along-track accuracy using four Doppler measurements.}
    \label{fig:alongtrack_startpoint}
\end{figure}

\subsection{Impact of the Observation Window Placement}
The previous subsection analyzed the inter-measurement interval while fixing the observation start time at the satellite rise epoch. In practice, however, the receiver may also choose where to place the observation window within a satellite pass. Since the satellite-receiver geometry changes rapidly during a LEO pass, the same number of measurements and the same inter-measurement interval may lead to different positioning performance depending on the window placement. This subsection therefore evaluates the effect of observation-window placement under multiple inter-measurement intervals.

\Cref{fig:crosstrack_startpoint} the cross-track positioning error obtained from four equally spaced Doppler measurements under different inter-measurement intervals. The horizontal axis denotes the time offset between the center of the observation window and the epoch of closest satellite-user approach, where zero corresponds to the minimum satellite-user range. The vertical axis denotes the cross-track error. For all tested intervals, the cross-track error exhibits a clear U-shaped trend with respect to the window-center offset. The minimum error is obtained when the observation window is centered near the closest approach. For a fixed inter-measurement interval, moving the window away from this point degrades the cross-track accuracy because the selected measurements become less balanced with respect to the satellite pass geometry. In addition, increasing the inter-measurement interval significantly reduces the cross-track error at the same window-center offset. With only four measurements, the cross-track error can be reduced from $10^5$ m to $10^3$ m just by increasing the observation interval time from 10 to 80 s. 

\Cref{fig:alongtrack_startpoint} shows along-track positioning errors obtained using four equally spaced Doppler observations from a single LEO satellite under different observation interval times. In contrast to the cross-track component, the along-track error exhibits a generally smaller magnitude and a weaker sensitivity to the observation timing offset. Although the along-track error decreases as the center of the observation window shifts, this change is less pronounced than in the cross-track direction. Furthermore, extending the observation interval time can improve the along-track error from approximately $600$ m to $100$ m. And the along-track error demonstrates smoother trends and smaller dynamic ranges compared to the cross-track error, suggesting that along-track direction is inherently more robust under limited-measurement conditions in single-satellite, multi-epoch LEO positioning.

The above results verified that both the inter-measurement interval and the temporal placement of the observation window play a critical role in single-satellite LEO Doppler positioning. Positioning the observation window's center point closer to the closest approach and employing a longer interval time significantly improves algorithm performance when few Doppler measurements are available. It can also be readily observed from \Cref{fig:crosstrack_startpoint} and \cref{fig:alongtrack_startpoint}, when the observation interval is set to 60 s, both cross-track and along-track errors can reach a relatively stable state, provided that the center of the selected observation window coincides with the epoch of closest satellite–user approach. 

\subsection{Positioning Error Evolution with Accumulated Observation Time}
\begin{figure}[t]
    \centering
    \includegraphics[width=1\linewidth]{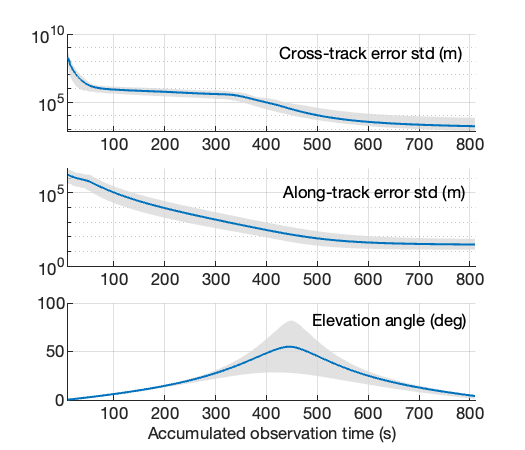}
    \caption{Evolution of positioning error with accumulated observation time during a single-satellite pass.}
    \label{fig:waitingtime}
\end{figure}

Complementary to the previous receiver-level parameter studies, this subsection examines a cumulative single-pass observation scenario. The receiver continuously collects Doppler measurements from a fixed start time, so increasing the accumulated observation time simultaneously increases the number of measurements and the observed satellite pass length.

\Cref{fig:waitingtime} shows the evolution of the cross-track and along-track positioning errors as a function of accumulated observation time. Both error components decrease as the observation time increases, indicating that extending the observation window improves the single-satellite solution through both measurement accumulation and the evolving satellite-user geometry. Quantitatively, the cross-track error decreases from approximately $10^8$ m at the beginning of the pass to the order of $10^4$ m after 350 s. The along-track error decreases from approximately $10^7$ m to below $10^2$ m over the same period. This confirms that extending the observation duration can significantly improve positioning performance during a single satellite pass.

The two error components exhibit different temporal behaviors. The along-track error decreases more smoothly and gradually approaches a stable level, which is consistent with the stronger Doppler sensitivity along the satellite motion direction. By contrast, the cross-track error remains relatively large during the early part of the pass and shows a more pronounced reduction around the middle of the observation interval. This indicates that the cross-track component is more strongly dependent on the geometric diversity accumulated over the pass. As a longer portion of the satellite trajectory is observed, the variation of the satellite-user line-of-sight geometry provides additional information for constraining the cross-track direction. After the main geometric variation has been captured, both error components tend to approach a stable level, suggesting that the additional benefit of further extending the observation window becomes limited.

To further distinguish the contribution of measurement accumulation from that of geometry evolution, the position gain is defined as $G_{pos}(t) = {P_0}/{P(t)}$, where $P_0$ is the error variance at first epoch ($M = 4$) and $P(t)$ is the error variance after accumulating observations up to time. The nominal gain caused only by the increasing number of measurements is modeled as $G_N(t) = {N(t)/N_0}$, following the $1/N$ error reduction law. The residual geometric gain is then defined as $G_{geo}(t) =G_{pos}(t)/ G_N(t)$. 

\begin{figure}[t]
    \centering
    \includegraphics[width=1\linewidth]{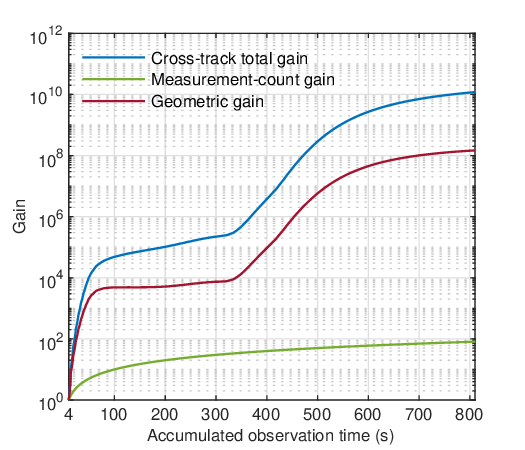}
    \caption{Measurement-Count and Geometric Contributions to Cross-track Error Reduction.}
    \label{fig:cross-factors}
\end{figure}

\begin{figure}[t]
    \centering
    \includegraphics[width=1\linewidth]{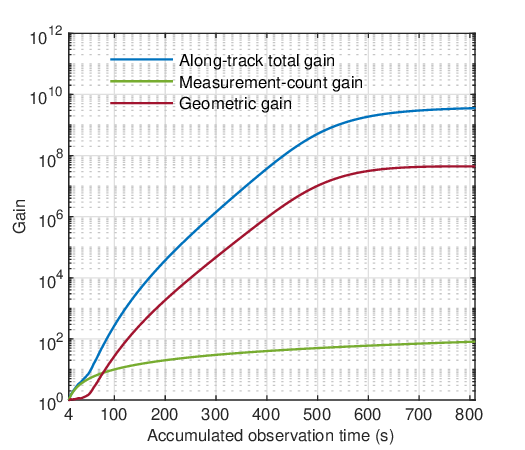}
    \caption{Measurement-Count and Geometric Contributions to Along-track Error Reduction.}
    \label{fig:along-factors}
\end{figure}

\Cref{fig:cross-factors} and \Cref{fig:along-factors} further separate the total position gain into the nominal measurement-count gain and the residual geometric gain for the cross-track and along-track directions, respectively. In both cases, the measurement-count gain increases only slowly with accumulated observation time, which indicates that the improvement expected from simply collecting more Doppler measurements is limited.

\Cref{fig:cross-factors} shows that the cross-track total gain is dominated by the residual geometric gain rather than the measurement-count gain. During the first approximately 50 s, both the total gain and the geometric gain increase rapidly, indicating that the initial segment of the pass provides useful geometric variation. From about 100 s to 330 s, the geometric gain remains nearly flat, and the total gain increases only slowly. During this interval, adding more Doppler measurements mainly provides limited measurement-count improvement and does not introduce significant new cross-track geometric diversity. After about 350 s, the geometric gain increases sharply, leading to a rapid increase in the total gain. This suggests that the main cross-track improvement occurs only after the satellite-receiver geometry has evolved sufficiently.

For the along-track component in \Cref{fig:along-factors}, unlike the cross-track case, the along-track geometric gain increases more continuously from about 100 s to 600 s, without a long flat region. After approximately 600 s, both the along-track positioning gain and the geometric gain begin to saturate, indicating that most of the useful along-track geometric information has already been accumulated. This confirms that the along-track direction benefits more gradually from the temporal evolution of the Doppler geometry, whereas the cross-track direction shows a delayed but sharper improvement.

Overall, this decomposition suggests that, for low-power IoT receivers, sparse Doppler measurements should be scheduled to capture informative geometry changes rather than simply increasing the number of closely spaced observations. This allows the receiver to reduce continuous tracking time while preserving most of the useful positioning information.

\section{Conclusion}

This paper investigated the geometric performance of single-satellite LEO Doppler positioning under sparse observation conditions. The DDOP-based framework is general for single-satellite multi-epoch Doppler positioning, while the numerical evaluation was carried out using Orbcomm TLE-propagated satellite trajectories as a representative LEO case. Therefore, the reported numerical values and specific thresholds should be interpreted within this Orbcomm-based scenario, whereas the observed geometric mechanisms provide broader insight into single-satellite LEO Doppler positioning.

The results show pronounced directional anisotropy in the positioning error. The cross-track direction is weakly constrained and can experience severe error amplification under unfavorable geometries, with errors reaching the order of $10^8$ m in the tested cases. In contrast, the along-track direction is more strongly constrained by Doppler measurements because it is closely related to the satellite motion direction. As a result, the along-track error remains much smaller and more stable across different satellite-user geometries.

The system-level geometric analysis shows that satellite-user geometry has a significant impact on positioning performance.
In terms of satellite pass maximum elevation, near-overhead passes improve along-track observability but degrade cross-track observability because the LOS vectors remain nearly coplanar. In the considered scenario, a maximum elevation of around $60^\circ$
 provides the best overall trade-off between along-track and cross-track errors. In addition, the user position relative to the satellite ground track strongly affects the cross-track error, with high-error regions aligned with the ground track. These results indicate that pass selection and geometry-aware operation are important for single-satellite LEO Doppler positioning.
 
The receiver-level analysis further provides several practical guidelines for sparse Doppler measurement scheduling. First, increasing the inter-measurement interval improves positioning performance significantly when the interval is short, while the improvement becomes limited when the interval approaches approximately 60 s in the considered scenario. Therefore, for constrained single-satellite Doppler positioning, an inter-measurement interval on the order of tens of seconds can provide a useful balance between geometric diversity and observation availability. Second, positioning accuracy strongly depends on the temporal placement of the observation window within the satellite pass. The best performance is obtained when the observation window is centered near the closest satellite-user approach, suggesting that low-power receivers should schedule their wake-up and measurement collection around the most informative portion of the pass. The accumulated-observation-time analysis further demonstrates that the benefit of a longer observation window is not solely due to the increased number of Doppler measurements. After removing the theoretical $1/\sqrt{N}$ measurement-count effect, the remaining geometric gain is especially pronounced in the cross-track direction, indicating that the observation window should cover the most geometrically informative portion of the satellite pass.

These findings highlight both the potential and the limitations of single-satellite LEO Doppler positioning under sparse observation conditions. The analysis provides a theoretical and numerical basis for geometry-aware pass selection, receiver-side measurement scheduling, and constrained IoT-oriented LEO-PNT operation. Future work will consider real-data validation, more realistic orbit and clock error models, and the extension of the analysis to other LEO satellites, multi-pass cases, and multi-satellite integration.

\bibliographystyle{IEEEtran}
\bibliography{ref}

\begin{appendices}
\section{Elements of the Jacobian matrix $\mathbf{H}$\label{appA}}
This appendix provides a detailed derivation of the measurement Jacobian matrix
\begin{eqnarray}
    \frac{\partial \dot \rho^j }{\partial   r_x}  &=&  \frac{-v_{s,x}^j}{\|\mathbf{r}-\mathbf{r}_s^j\|} \notag \\&+& \frac{(\mathbf{v}_s^j)^T (\mathbf{r}-\mathbf{r}_s^j)(r_x - r_{s,x}^j)}{\|\mathbf{r}-\mathbf{r}_s^j\|^3}\\
    \frac{\partial \dot \rho^j }{\partial r_y}  &=&  \frac{-v_{s,y}^j}{\|\mathbf{r}-\mathbf{r}_s^j\|^3} \notag \\&+& \frac{(\mathbf{v}_s^j)^T (\mathbf{r}-\mathbf{r}_s^j)(r_y - r_{s,y}^j)}{\|\mathbf{r}-\mathbf{r}_s^j\|^3} \\
    \frac{\partial \dot \rho^j }{\partial r_z}  &=&  \frac{-v_{s,z}^j}{\|\mathbf{r}-\mathbf{r}_s^j\|^3} \notag \\&+& \frac{(\mathbf{v}_s^j)^T (\mathbf{r}-\mathbf{r}_s^j)(r_z - r_{s,z}^j)}{\|\mathbf{r}-\mathbf{r}_s^j\|^3}\\ 
    \frac{\partial \dot \rho^j }{\partial \dot\delta_d}  &=& c \\
    \frac{\partial \dot \rho^j }{\partial \delta _t^j} & =&(\mathbf{a}_s^j)^T\frac{(\mathbf{r}-\mathbf{r}_s^j)}{\|\mathbf{r}-\mathbf{r}_s^j\|} - \frac{(\mathbf{v}_s^j)^T\mathbf{v}_s^j}{\|\mathbf{r}-\mathbf{r}_s^j\|} \notag \\&+&\: \frac{(\mathbf{v}_s^j)^T(\mathbf{r}-\mathbf{r}_s^j)(\mathbf{r}-\mathbf{r}_s^j)^T\mathbf{v}_s^j}{\|\mathbf{r}-\mathbf{r}_s^j\|^3}
\end{eqnarray}
where the user position and satellite position are $\mathbf{r} = [r_x,r_y,r_z]$,$\mathbf{r}_s^j = [r_{s,x}^j,r_{s,y}^j,r_{s,z}^j]$, the satellite velocity is $\mathbf{v}_s^j = [v_{s,x}^j,v_{s,y}^j,v_{s,z}^j]$ and $\mathbf{a}_s^j$ is the satellite acceleration.
\end{appendices}

\end{document}